\documentclass[trackchanges, twocolumn]{aastex701}

\usepackage{amsmath}
\usepackage{textcomp}
\usepackage{xcolor}
\usepackage{placeins}
\shorttitle{The JWST Search for Earth--Luna Analogs: TOI 700}
\shortauthors{Pass et al.}
\received{April 6, 2026}
\revised{July 8, 2026}
\accepted{July 28, 2026}
\submitjournal{AJ}
\begin{document}

\title{The JWST Search for Earth--Luna Analogs: \\Upper Limits on Exomoons and Refined Ephemerides for TOI 700 d and e}

\author[0000-0002-1533-9029]{Emily K. Pass}
\affiliation{Kavli Institute for Astrophysics and Space Research, Massachusetts Institute of Technology, Cambridge, MA 02139, USA}
\email{epass@mit.edu}

\author[0000-0002-9003-484X]{David Charbonneau}
\affiliation{Center for Astrophysics $\vert$ Harvard \& Smithsonian, 60 Garden Street, Cambridge, MA 02138, USA}
\email{}

\author[0000-0001-7246-5438]{Andrew Vanderburg}
\affiliation{Center for Astrophysics $\vert$ Harvard \& Smithsonian, 60 Garden Street, Cambridge, MA 02138, USA}
\email{}

\author[0000-0003-4733-6532]{Jacob L. Bean}
\affiliation{Department of Astronomy \& Astrophysics, University of Chicago, Chicago, IL 60637, USA}
\email{}

\begin{abstract}
\noindent While no conclusive detections of exomoons have been reported to date, planet formation theories predict that Luna-analog satellites should be a common outcome of the collisional dynamics in early extrasolar systems. Such satellites have the potential to unlock new avenues to learn about exoplanet systems, speaking to topics of habitability, tidal heating, planet formation, late-stage growth, planetary compositions, and more. Here we describe the results of our JWST program to search for Luna-analog exomoons around the rocky, habitable-zone M-dwarf planets TOI 700 d and e. We refine the ephemerides of both worlds, providing an order-of-magnitude improvement in period precision and a factor of 2--3 improvement in planetary radii. We identify a strong correlated noise signal with a timescale of $16\pm4$ minutes and an amplitude of $46\pm4$ ppm; similar signals have been observed in previous JWST analyses of other stars and have been ascribed to stellar granulation. This noise source inflates our error by a factor of 4 relative to photon-noise expectations in 10-minute bins and limits our sensitivity to moons: we determine that our observations are sensitive mainly to moons larger than Ganymede on periods longer than 2 days (i.e., moons larger than our solar system's natural satellites). If this noise could be corrected, we would be sensitive to Luna-analog moons. Future work to address this noise source will thus be critical for detecting exomoons in stellar transits, as well as for all other science cases that hope to take advantage of JWST white-light curves in the photon-noise limit. 

\end{abstract}

%\keywords{\uat{Galaxies}{573} --- \uat{Cosmology}{343} --- \uat{High Energy astrophysics}{739} --- \uat{Interstellar medium}{847} --- \uat{Stellar astronomy}{1583} --- \uat{Solar physics}{1476}}


\section{Introduction} 
Our moon plays an important role for our planet. It is thought that the presence of Luna stabilizes Earth's obliquity, preventing chaotic variations that would have destabilized Earth's climate \citep{Laskar1993}. In this sense, large moons may be important for habitability; the obliquity of Mars, in contrast, remains chaotic to this day \citep{LaskarRobutel1993}. Tidal heating of the Earth due to Luna may also have been an important heat source for the early Earth, affecting the thermal and compositional evolution of the mantle \citep{Zahnle2007}. Some authors have even argued that these early tidal effects were needed for the onset of prebiotic \hbox{chemistry, leading to the origin of life \citep{Lathe2004}.}

Given the importance of Luna to our planet, it is natural to wonder: how common are similar satellites around exoplanets? Based on the four terrestrial planets of the solar system, one might naively estimate 25\%; alternatively, numerical simulations by \citet{Elser2011} found rates as high as 1 in 4 or as rare as 1 in 45. To date, extrasolar observations have not been sensitive enough to sharpen this picture, and so the rarity of Luna analogs remains unclear.

While the study of exomoons is in its infancy, these satellites ultimately hold the promise to inform a wide array of intriguing science questions:

\textit{Planet formation--} The moons of small planets are thought to be formed from planet-scale collisions; the demographics of such moons (in size, mass, and location) can therefore provide insight into a stellar system's evolution in  the late stages of planetesimal accretion, informing planet formation models (e.g., \citealt{Barr2016}).

\textit{Moon systems--} While our assumption from these models is that terrestrial planets will host singleton moons, it could be that we will find systems of moons, suggesting an in-situ formation mechanism akin to the Galilean satellites; in our solar system, study of the composition gradient of the Galilean satellites has led to an understanding of the temperature environment in which the satellites formed and has placed constraints on the late-stage growth of gas giants \citep{Pollack1974, Canup2002}.

\textit{Planet properties--} By studying the orbit of a moon, one can measure the planetary density, allowing for an independent determination of the planetary mass, even in cases where it cannot be constrained by radial velocities \citep{Kipping2010}. One can also measure the planet's tidal dissipation factor and constrain the interior properties of the planet, including distinguishing between rocky and volatile-rich compositions (e.g., \citealt{Tokadjian2022}). For systems for which the moon and planet undergo mutual eclipses, one could use high-quality photometry and the technique of eclipse mapping to resolve the surface features on the planet \citep{Buie1992}.

\textit{Rings--} In addition to moons, one can search for circumplanetary rings, like the rings of Saturn, which induce other detectable distortions to the transit light curve \citep{Barnes2004, Ohta2009, Tusnski2011}; such rings may indicate the recent destruction of a moon that has ventured inside the Roche limit.

While these lines of inquiry are exciting, they are also premature: before we can study exomoons, we must first find them. Past searches, such as the Hunt for Exomoons with Kepler \citep[HEK;][]{Kipping2012}, have focused on very large moons ($>$10$^{-1}$M$_\oplus$) around gas giants, a type of satellite that does not exist in our solar system. This strategy has been necessary due to Kepler's lack of sensitivity to smaller satellites.\footnote{That said, HEK also placed a constraint on the occurrence of Galilean-analog moon systems at a population level by stacking phase-folded transits of many planets \citep[][$<$0.38 at 95\% confidence for planets between 0.1--1.0~AU]{Teachey2018}.} While no conclusive detections have yet emerged, two candidates have been presented in the form of Kepler~1625~b-i \citep{Teachey2018} and Kepler~1708~b-i (\citealt{Kipping2022}; but see counterarguments in \citealt{Heller2024}). Both of these candidates are Neptune-sized, and therefore completely unlike any of the natural satellites of our solar system. Work is underway to use the James Webb Space Telescope \citep[JWST;][]{Gardner2006} to push down to the sensitivity necessary to detect Ganymede analogs around Jupiter-like planets, although a pilot study of Kepler~167~e revealed that such searches are currently limited by long-term trends in JWST NIRSPEC data \citep{Kipping2026}. Another group is using JWST to search for Galilean-analog satellites around free-floating giant planets \hbox{\citep{Wilson2025}}. Others are using VLTI/GRAVITY in hopes of detecting exceptionally large moons around giant planets through astrometry \citep{Winterhalder2026, Kral2026, Macias2026}, and simulations indicate that the upcoming Roman Space Telescope may discover a small number of Ganymede-analog or larger moons through microlensing \citep{Lastovka2025}.

Here we focus on a different type of moon: a Luna-sized satellite that orbits a terrestrial planet, i.e., a true analog of our own planet-moon system. The transit of Luna across our Sun would induce a dip of just 6 ppm -- a tiny signal, impossible to detect with existing instrumentation. However, the depth of the transit scales with the square of the stellar radius. If you lower the radius of the star to 0.56R$_\odot$, an early M dwarf, a Luna-sized satellite would induce a 20 ppm dip, with the signal growing to hundreds of ppm as you lower the radius into the mid-to-late M dwarf regime. JWST is capable of detecting such signals: \citet{Coulombe2023} found that the observations of WASP~18 with NIRISS/SOSS from the Early Release Science Program achieved RMS errors of 5 ppm in 1-hr bins. By studying M-dwarf planets around which a moon could be dynamically stable, JWST could plausibly achieve the first conclusive detection of an exomoon.

In this paper, we report first results from such a search for Earth--Luna analogs with JWST. In Section~\ref{sec:data}, we describe our target selection and data collection. In Section~\ref{sec:analysis}, we present our analysis: the JWST observations alone in Section~\ref{sec:jw}, jointly with TESS/Spitzer data in Section~\ref{sec:joint}, and our search for exomoons in Section~\ref{sec:search}. We conclude in Section~\ref{sec:conclusion}.

\section{Observations}
\label{sec:data}
\subsection{Target selection}
\label{sec:selection}
\begin{figure}[t]
\centering
    \includegraphics[width=\columnwidth, trim=3 5 3 5,clip]{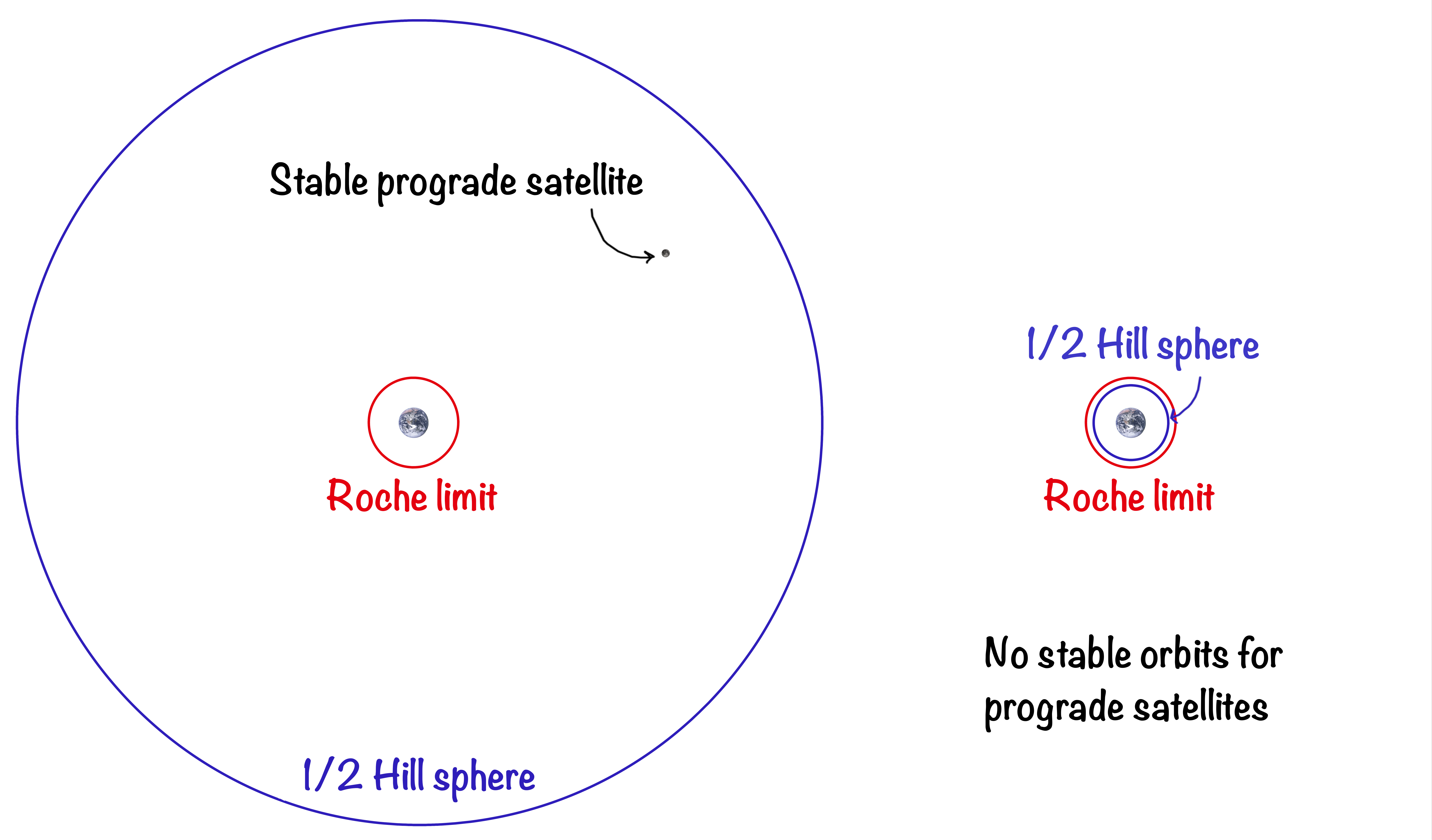}
    \caption{\textbf{Left:} Stable orbits are possible for prograde moons located outside the planet’s Roche limit and inside half the Hill sphere. This diagram has been drawn to scale for a Luna-sized satellite orbiting TOI 700 d. \textbf{Right:} If the Roche limit is larger than half the Hill sphere, no prograde moons would be stable.}
    \label{fig:diagram}
\end{figure}

To identify the best candidates to search for exomoons, we consider the Transiting Exoplanet Survey Satellite \citep[TESS;][]{Ricker2015} Objects of Interest catalog \citep{Guerrero2021} as of 11 October 2023, and crossmatch with the TESS Input Catalog \citep{Stassun2019} to obtain 2MASS magnitudes and stellar masses. We then estimate the planetary mass based on the observed radius. For planets smaller than 1.7R$_\oplus$, we use the terrestrial mass--radius relation from \citet{Zeng2019}. For more massive planets, we use the mass--radius relation from \citet{Chen2017}, capping the mass at 1M$_{\rm J}$ as this relationship is degenerate in the Jovian-radius regime. For each planet, we consider whether there is a region where a Luna-sized (0.27R$_\oplus$) satellite could exist in a stable orbit (Figure~\ref{fig:diagram}); i.e., whether the Roche limit, where the planet would tear apart the moon, is smaller than the maximum separation a moon could have before the star would tear it away from the planet (0.4895 of the Hill radius for a prograde satellite;  \citealt{Domingos2006}). While the dynamical influence of other planets can slightly reduce this maximum separation, the reduction is expected to be modest, even in closely packed systems \citep{Payne2013}.

Due to these restrictions, whether or not a planet is capable of hosting a dynamically stable moon is dependent on the mass and radius of the planet, the mass of the star, and the orbital period of the planet. Moon orbits also evolve over time, although there is debate surrounding the influence of tidal interactions between a star, its tidally locked planet, and its moon on limiting the range of dynamically allowed orbits. While \citet{Barnes2002} calculated that such effects could be significant in limiting moon lifetimes (a concern echoed by \citealt{Patel2026}), recent work by \citet{Kisare2024} showed that tidal dissipation within the satellite can prevent Hill sphere escape, leading to a much larger range of moons that are stable on long timescales. Empirical constraints on exomoon occurrence rates are thus necessary to test whether tidally locked planets can retain satellites like our Luna.

\begin{figure*}[t]
\centering
    \includegraphics[width=0.74\textwidth]{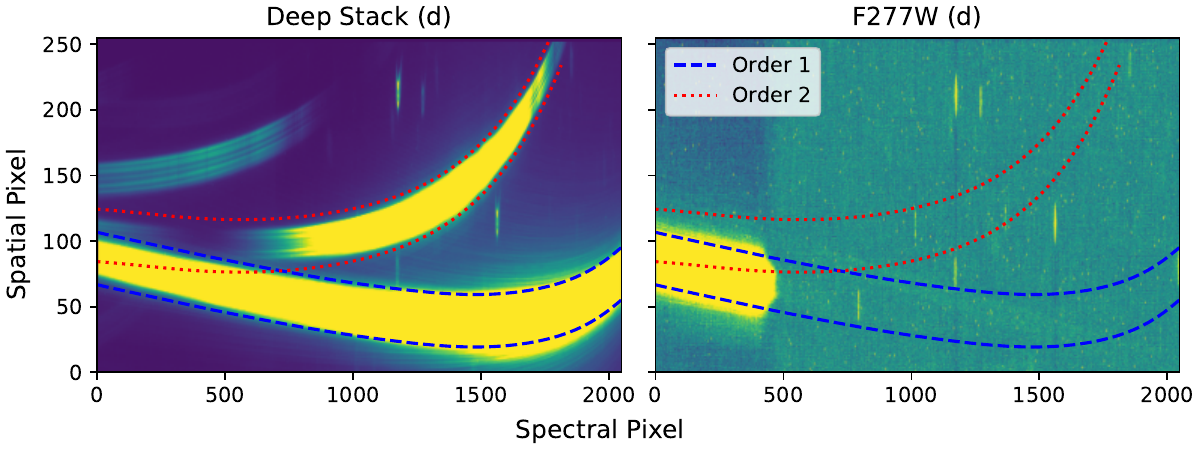}
    \includegraphics[width=0.74\textwidth]{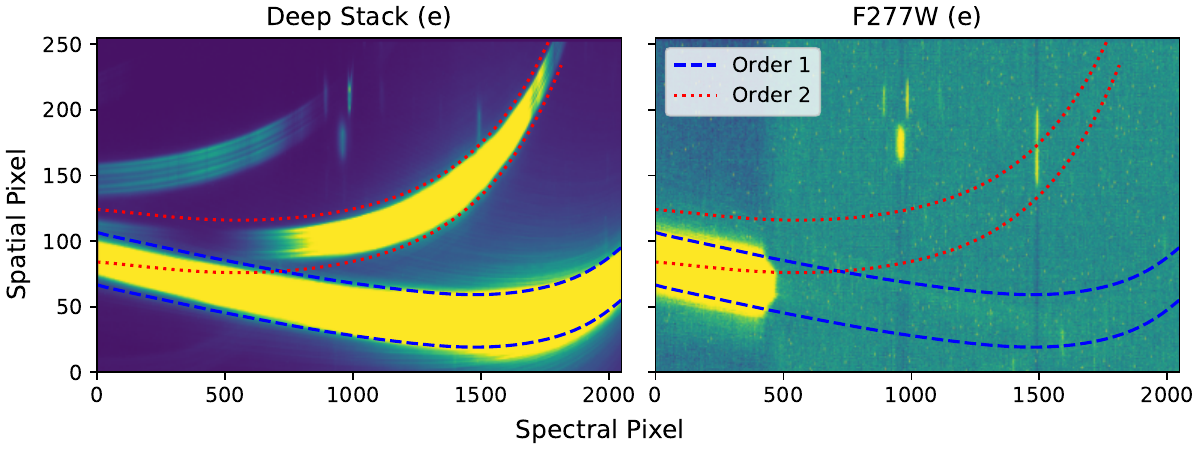}
    \caption{\textbf{Left:} A median combination of our observations of TOI 700 d (upper panel) and e (lower panel). The dashed/dotted lines show our extraction apertures. \textbf{Right:} An observation taken with the F277W filter. As only the reddest end of order 1 is transmitted, this filter reveals background contaminants. We carefully selected our aperture position angle constraints so that only a few contaminants are present within our apertures; we mask these wavelengths before producing our white-light curves.}
    \label{fig:f277w}
\end{figure*}

As we are interested in moons that could be detectable with JWST, we further limit our target selection to systems where the transit of a satellite with the same radius as Luna would have a transit depth exceeding 20 ppm and the SNR of the moon transit is $>$5$\sigma$. For the purposes of ranking potential targets, we make an initial estimate of the SNR using the empirical noise in 10-minute bins for WASP~18 as observed by NIRISS/SOSS \citep{Coulombe2023} scaled photometrically based on 2MASS $J$-band magnitude. We include a noise floor at 5 ppm, as JWST's performance has not been proven beyond that level. After selecting our candidates, we refine our SNR estimates using the JWST Exposure Time Calculator and our specific observing strategy.

This selection reveals that the TOI 700 system is singularly suited for our purposes: if we sum the phase-space area in which moons would be dynamically stable for all potential targets, TOI 700~d and e together form nearly half the total. These worlds are roughly Earth-sized planets orbiting in the habitable zone of their mid-M dwarf star \citep{Gilbert2020, Gilbert2023}. Since JWST does not have sufficient precision to study the atmospheres of these planets \citep{Suissa2020}, they are unlikely to be otherwise targeted by the telescope, motivating us to propose and be awarded time to observe these planets under program GO 6193 (PI:~Pass).

While TOI 700~d and e are by far the best targets to search for Luna-analog moons,\footnote{Throughout this work, we use `Luna analog' to refer to moons roughly the size of Luna orbiting small planets. Given the more compact nature of M-dwarf systems, these satellites would necessarily orbit closer to their planets than does Luna; specifically, the Earth-Luna distance is roughly double the maximum dynamically allowed separation for a satellite of TOI 700 d.} other planets do have some dynamical space in which moons may exist, including some with archival observations from JWST. In an upcoming companion paper, we will apply the methods from this work to these archival data sets to produce population-level occurrence rate constraints.

\subsection{Data collection and reduction}
\label{sec:toi700}

We observed one transit of each of planets d and e with JWST using the Near Infrared Imager and Slitless Spectrograph (NIRISS; \citealt{Doyon2023}) in Single Object Slitless Spectroscopy (SOSS; \citealt{Albert2023}) mode with the SUBSTRIP256 subarray. The observations were taken on 28 March 2025 for d (9.0 hours) and 6 April 2025 for e (8.3 hours) using 4 groups per integration, resulting in 1181 integrations of d and 1087 integrations of e. The transit durations of d and e are 3.3 hours and 3.1 hours, respectively; our observations include sufficiently long out-of-transit baseline to ensure that the transit of a satellite orbiting at the maximum dynamically allowed separation (roughly half the Hill radius; \citealt{Domingos2006}) would not be missed.

We reduce the JWST observations using the \texttt{exoTEDRF}\footnote{Previously known as \texttt{supreme-SPOON}} pipeline \citep{Radica2024}, which has been widely used for NIRISS/SOSS observations \citep[e.g.,][]{Feinstein2023, Radica2023, Radica2024b, Radica2025, Coulombe2023, Lim2023, Cadieux2024, Fournier-Tondreau2024, Piaulet2024, Taylor2025}. Our reduction procedure is similar to the processes outlined in those works, comprising Stages 1--3 of the \texttt{exoTEDRF} workflow. To summarize, Stage 1 includes superbias, group-level 1/$f$, and non-linearity corrections, as well as cosmic-ray flagging and ramp fitting; Stage 2 includes flat fielding, background subtraction, and bad pixel correction; Stage 3 includes order tracing and spectral extraction using a simple box aperture with a width of 40, the size of which we selected to minimize the noise in the resulting white-light curve. During the 1/$f$ correction, \texttt{exoTEDRF} masks a number of features to avoid biasing the noise estimation: background contaminants using a F277W exposure (Figure~\ref{fig:f277w}), the detector reset artifact, pixels that deviate by more than 10$\sigma$, the spectral traces themselves, and any pixel with a non-zero data quality (DQ) flag. We include the first four of these masks, but diverge slightly from the standard reduction by neglecting the DQ mask, as we find that this masking slightly increases the noise in our output white-light curves. This change only affects the 1/$f$ correction; later in the workflow, pixels with DQ flags are still corrected using interpolation. 

\begin{deluxetable}{cr}[t]\label{tab:wlc}
\tablecaption{NIRISS/SOSS white-light curve of TOI 700}
\tablewidth{300pt}
\tablecolumns{2}
\tablehead{ 
\colhead{Time (BJD [TDB] $-$ 2400000.5)} & 
\colhead{\phantom{extra} $\Delta$Flux (ppm)}} 
\startdata 
60762.32561 & $-$199\\
60762.32593 & $-$156\\
60762.32625 & $-$186\\
60762.32657 & 59\\
60762.32689 & $-$66\enddata
\tablecomments{Full table available in machine-readable format.}
\end{deluxetable}

\begin{figure*}[h]
\centering
    \includegraphics[width=0.49\textwidth]{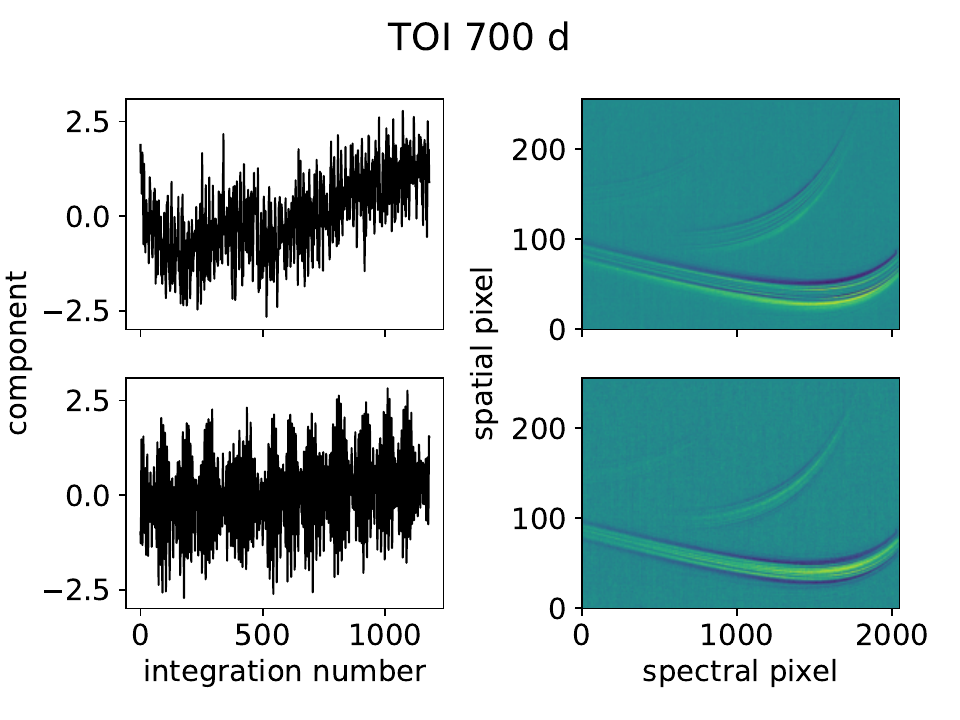}
    \includegraphics[width=0.49\textwidth]{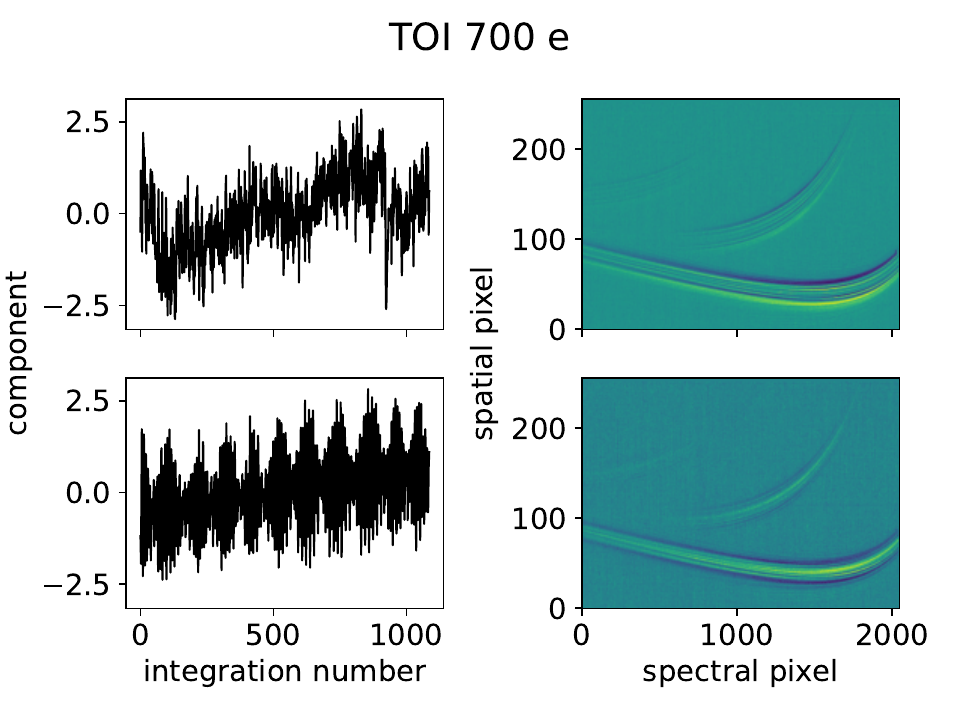}
    \caption{Morphological changes of the spectral trace during our observations, as determined from principal-component analysis (PCA) by following a similar method to \citet{Coulombe2023}. We use the \texttt{scikit-learn} \citep{Pedregosa2011} routine \texttt{IncrementalPCA}, applying it to the 3D data cube after normalization using \texttt{scikit-learn}'s \texttt{StandardScaler}. We find that two PCA terms are sufficient to describe the systematic trends, and our identified signals are similar to those presented in \citet{Coulombe2023}: a trend associated with changes to the y-position of the trace (upper panels) and a beat pattern associated with changes to the FWHM of the trace (lower panels). This latter effect has been attributed \citep[e.g.,][]{Piaulet2024} to thermal oscillations in the telescope \citep{McElwain2023}.}
    \label{fig:pca}
    \vspace{1cm}
    \centering
    \includegraphics[width=1.05\columnwidth]{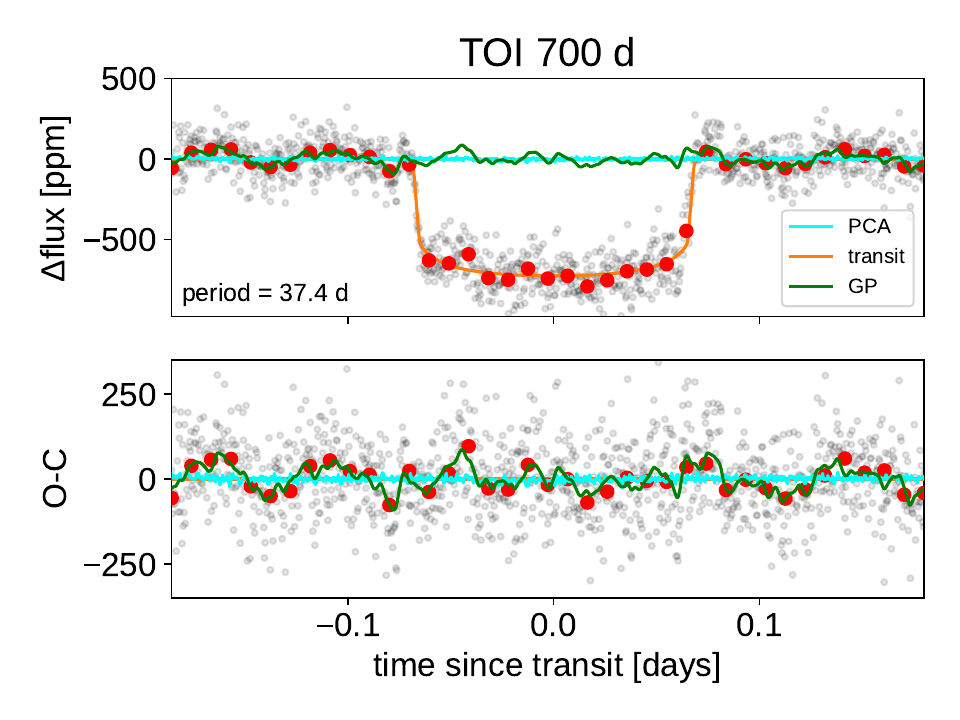}
    \includegraphics[width=1.05\columnwidth]{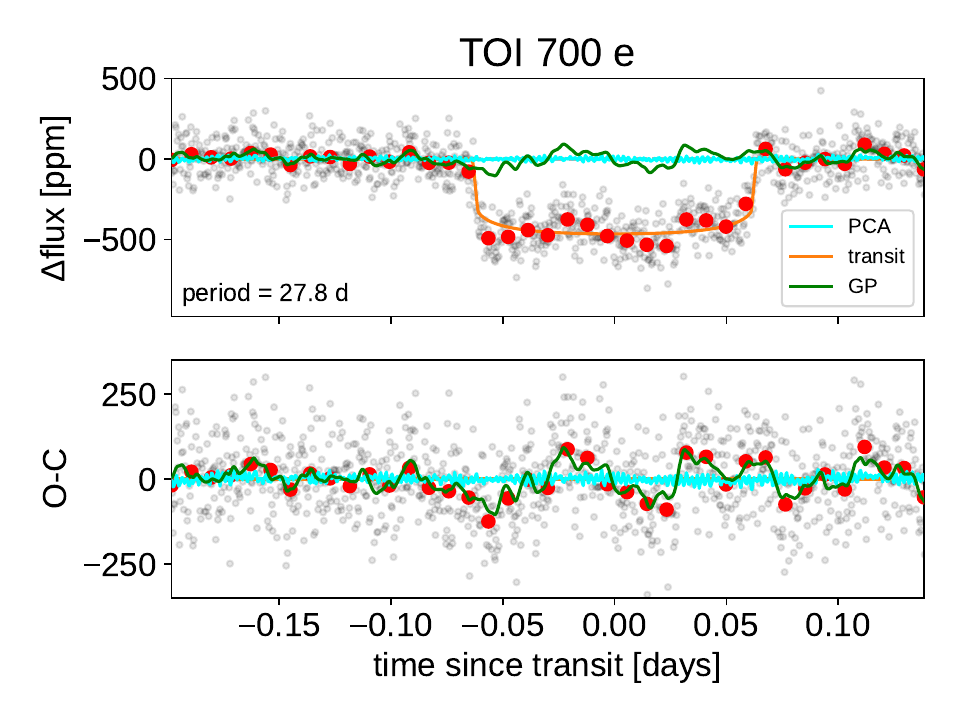}
    \caption{Our NIRISS/SOSS transits of TOI 700 d and e. Unbinned data are in gray, binned data in red, the best-fit transit model in orange, and the systematics model in blue. The lower panels show the residuals after subtracting the transit model. Our PCA-determined instrumental systemics model is unable to explain the correlated noise in the residuals; we quantify this excess correlated noise by fitting a Gaussian process (GP), shown in green. We use the same GP hyperparameters for the two transits, with a best-fitting timescale of 16$\pm$4~minutes and amplitude of 46$\pm$4~ppm. Without a correction, the RMS of the residuals in 10-minute bins is 39~ppm for d and 44~ppm for e, or 12~ppm for both if we apply the GP correction.}
    \label{fig:transits}
\end{figure*}

After processing our data through Stages 1--3 of \texttt{exoTEDRF}, we have a data cube for each planet with dimensions of time, wavelength, and flux. To produce a white-light curve, we sum across all wavelengths of order 1 (0.85--2.83\textmu m) and 0.6--1.0\textmu m for order 2 (i.e., neglecting regions of low counts). We also mask wavelength regions where we identified a bright background contaminant in our F277W exposures: for TOI~700~d, this is $\lambda < 0.86$\textmu m and 2.044--2.053\textmu m in order 1; for TOI~700~e, this is 0.705--0.721\textmu m in order 2. Our resulting white-light curve is given in Table~\ref{tab:wlc}, normalized to the out-of-transit baseline.

\section{Fitting and Analysis}
\label{sec:analysis}
In Section~\ref{sec:jw}, we consider the JWST data in isolation, describing and justifying the various choices we make in the fitting process. In Section~\ref{sec:joint}, we combine these data with archival Spitzer and TESS photometry to determine the best-fit orbital parameters for the system. In Section~\ref{sec:search}, we search for moons, comparing the planet-only and planet+moon fits and conducting an injection and recovery analysis to evaluate our sensitivity. 

\subsection{JWST-only fit}
\label{sec:jw}

\begin{figure*}[h]
\centering
    \includegraphics[width=1.\textwidth]{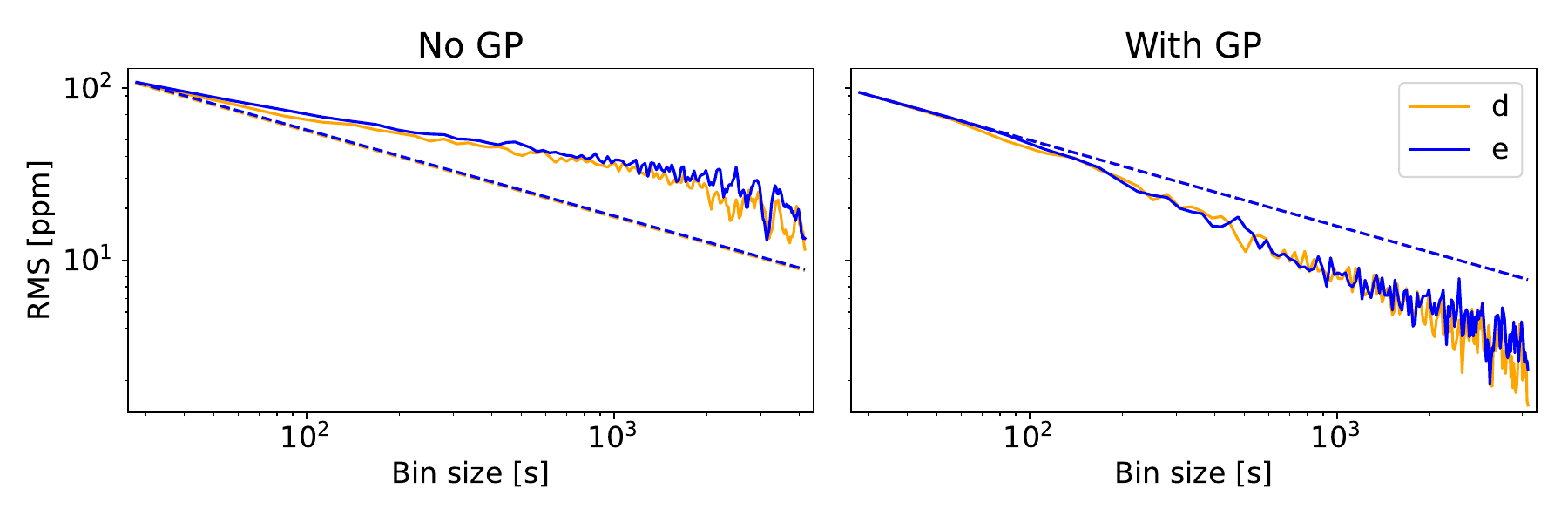}
    \caption{RMS residuals of our light curve fits at different binning cadences. We analyze the two transits separately but the results are comparable, with marginally worse performance for e. The dashed lines are a prediction of RMS improvement following an empirical white-noise scaling (i.e., the native-cadence RMS scaled by $1/\sqrt{N}$, where $N$ is the number of points per bin). Without a Gaussian process (GP; left), the binning performs worse than the prediction due to the presence of time-correlated noise (potentially stellar granulation; see text). Unlike the findings of \citet{Coulombe2023} for WASP~18~b, we are still unable to recover the white-noise scaling after binning down to cadences longer than a hour. With a GP (right; described in Section~\ref{sec:stellar}), we achieve better precision than we expect from this scaling. While this could indicate over-fitting on the part of the GP, it could alternatively suggest that there is also excess noise at high frequencies; therefore, scaling from the achieved RMS at the native 27s sampling cadence would underestimate the achievable precision. We note that if we instead adopt the nominal flux errors from \texttt{exoTEDRF} to calculate our expected RMS at the native cadence and scale from this value, the GP model provides excellent agreement with Poisson expectations when binned to \hbox{cadences equal or longer than the GP length scale.}}
    \label{fig:binning}
    \vspace{1cm}
    \centering
    \includegraphics[width=1.0\columnwidth]{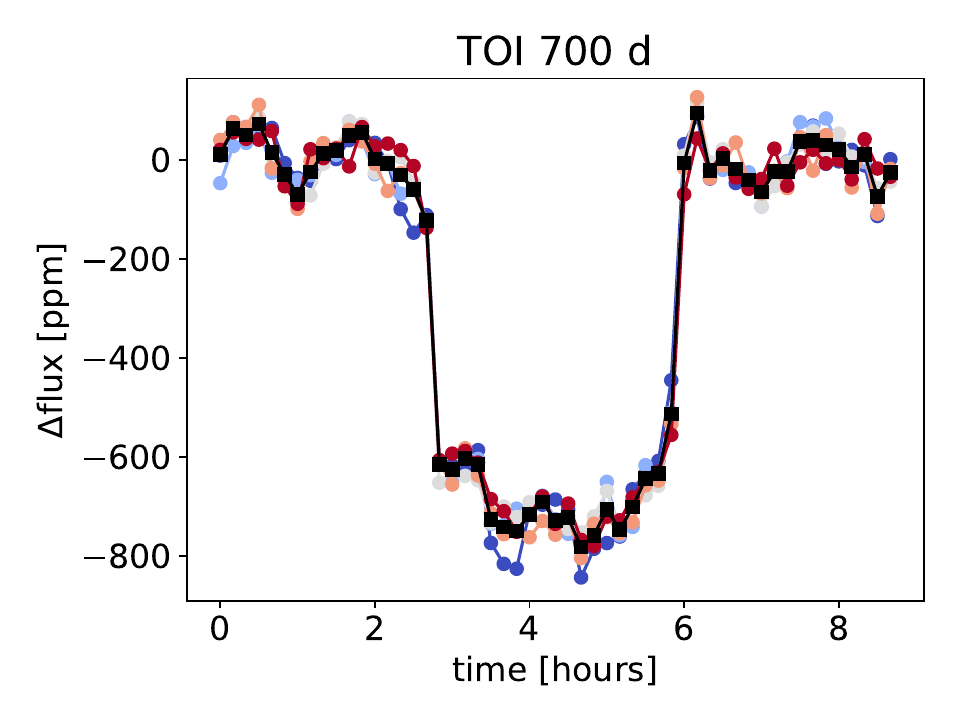}
    \includegraphics[width=1.0\columnwidth]{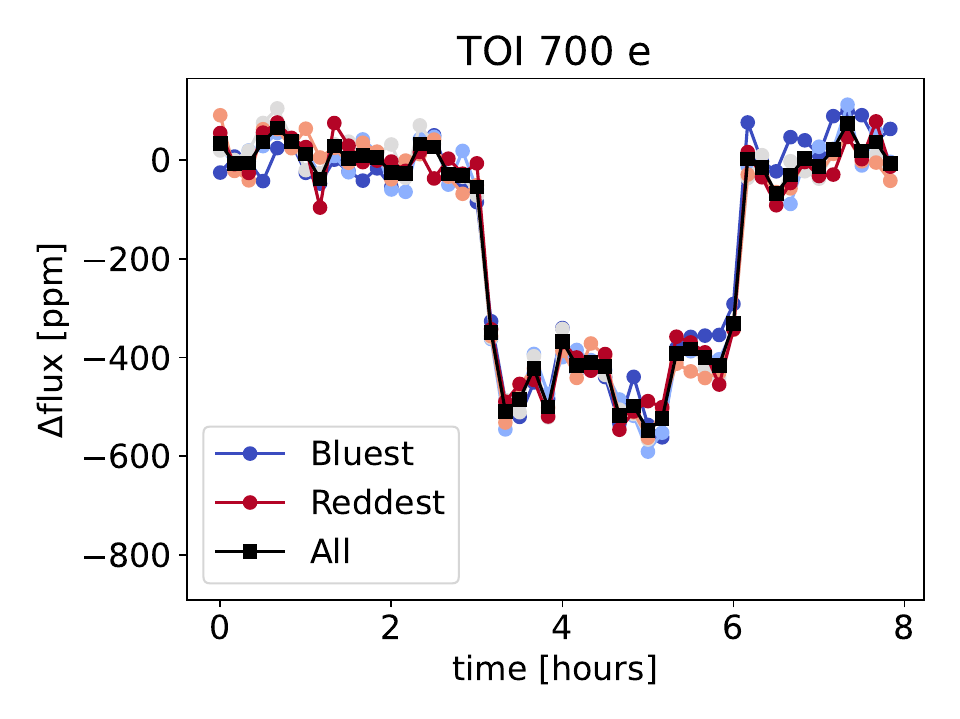}
    \caption{We subdivide our light-curve data into five wavelength bins, each contributing 20\% of the total flux. In terms of wavelengths, these bins correspond to 0.60--1.02 (labeled `bluest'), 1.02--1.21, 1.21--1.42, 1.42--1.74, and 1.74--2.83\textmu m (labeled `reddest'). The data are also binned to 10-minute intervals in the time axis. In black, we show our white-light curve that includes all wavelengths. Much of the correlated noise appears to be consistent across all wavelengths. Nonetheless, we do observe some chromatic systematics, particularly for the bluest bin, which shows a slope during the transit of TOI~700~e and stronger correlated noise during the transit of TOI~700~d.}
    \label{fig:chromatic}
\end{figure*}

\subsubsection{Modeling framework}
We use the \texttt{exoplanet} package \citep{ForemanMackey2021} to model and fit our system. As we are considering only the JWST data in this section, we do not include planet b or c in this model, and we constrain the periods of planets d and e using a prior informed by the joint fit that we will describe in Section~\ref{sec:joint}. We adopt the stellar mass and radius measurements from \citet[][0.419$\pm$0.021R$_\odot$ and 0.417$\pm$0.021M$_\odot$]{Gilbert2023}, which they estimated using the absolute $K_s$-band relations from \citet{Mann2015}. We use \texttt{exoCTK} \citep{Bourque2021} to estimate quadratic limb-darkening parameters for NIRISS/SOSS white-light curves using the \texttt{PHOENIX ACES} stellar model grid \citep{Husser2013}, yielding $u_1$=0.113$\pm$0.011 and $u_2$=0.223$\pm$0.017. We conservatively inflate the errors in the limb-darkening priors to $\pm$0.1 to mitigate the known biases and model-dependent uncertainties intrinsic to predicting quadratic limb-darkening parameters from model atmospheres (e.g., \citealt{Espinoza2015}). We assume circular orbits (although this assumption will be relaxed in Section~\ref{sec:ecc}) and adopt uninformative uniform priors for the other orbital parameters. For each transit, we also fit for a normalization term, and \hbox{a white-noise jitter term to inflate our observed errors.}

\subsubsection{Instrumental systematics}
Other works analyzing NIRISS/SOSS observations correct for systematics associated with morphological changes to the spectral trace \citep[e.g.,][]{Coulombe2023}, including linear terms in their white-light curve fit with respect to eigenvectors from a principal-component (PCA) analysis. We perform such an analysis for our TOI 700 data, with the resulting eigenvectors and eigenimages shown in Figure~\ref{fig:pca}. This investigation reveals two morphological changes that affect the trace during our observations: a change in y-axis position and a change in FWHM, similar to the systematics identified by \citet{Coulombe2023} in their analysis of WASP~18~b. We investigate detrending linearly against these terms in our white-light curve fit, but ultimately find that this treatment has a negligible impact on our results: for TOI~700~d, the RMS in 10-minute bins is unchanged, while TOI~700~e improves from 44~pm to 43~pm (Figure~\ref{fig:transits}). We also note that a Lomb-Scargle analysis of our light curves does not reveal any significant power at the periods associated with the beat pattern from the PCA analysis, further supporting that these instrumental systematics are not large enough to substantially affect the light curve. Instead, the residuals are dominated by correlated noise that occurs on a longer length scale, which previous works have attributed to stellar granulation \citep[][]{Cadieux2024, Coulombe2025}. This correlated noise dominates the error budget, even after binning down to one-hour timescales (Figure~\ref{fig:binning}). We investigate this noise in the following section.

\subsubsection{Stellar systematics}
\label{sec:stellar}
Plugging the stellar parameters for TOI 700 into the scaling relations for granulation from \citet{Gilliland2011}, we find
\begin{multline}
\tau_{\rm gran} = \\ 220 \Big ( \frac{0.417M_\odot}{M_\odot} \Big)^{-1}\Big(\frac{0.419R_\odot}{R_\odot}\Big)^2\Big(\frac{3459\rm{K}}{T_{\rm{eff},\odot}}\Big)^{0.5}~\rm{s} \\
= 73\rm{s}
\end{multline}
and
\begin{multline}
\sigma_{\rm gran} = \\
75 \Big ( \frac{0.417M_\odot}{M_\odot} \Big)^{-0.5}\Big(\frac{0.419R_\odot}{R_\odot}\Big)\Big(\frac{3459\rm{K}}{T_{\rm{eff},\odot}}\Big)^{0.25}\rm{ppm} \\
= 43~\rm{ppm}.
\end{multline}
\citet{Gilliland2011} infer these scaling relations from the asteroseismic relation $\nu_{\rm max} \propto MR^{-2}T_{\rm eff}^{-0.5}$ \hbox{\citep{Kjeldsen1995}} and the assumptions that $\sigma_{\rm gran} \propto \nu_{\rm max}^{-0.5}$ \citep{Chaplin2011} and $\tau_{\rm gran} \propto \nu_{\rm max}^{-1}$ \citep{Kjeldsen2011}. That said, such oscillations have yet to be detected in an M dwarf, and so the use of these relations is an extrapolation in the M-dwarf regime. Furthermore, these relations are calibrated for Kepler; we would expect the granulation amplitudes to be smaller in the redder NIRISS/SOSS band. However, the dependency on wavelength may be complex: in their simulations of stellar granulation using the \texttt{STAGGER} grid of 3D radiative hydrodynamical models \citep{Magic2013}, \citet{Chiavassa2017} found that while the amplitude of granulation-induced variations can be as much as a factor of three greater in the optical than the infrared for a Sun-like star, the fluctuations among different wavelengths are stronger in the infrared than in the visible. In our observations, we find that the bluest wavelengths show somewhat more correlated noise than redder wavelengths, although there is not a consistent correlation between wavelength and amplitude (Figure~\ref{fig:chromatic}).

To assess whether our correlated noise is consistent with expectations for stellar granulation, we add a Gaussian process (GP) with a simple harmonic oscillator kernel to our fit. Specifically, we use the \texttt{SHOTerm} kernel as implemented in \texttt{celerite2} \citep{ForemanMackey2018}, with quality factor $Q$=1/3 and fitting for the undamped period and standard deviation. We find that the red noise has a characteristic timescale of 16$\pm$4~minutes and a characteristic amplitude of 46$\pm$4~ppm (shown in green in Figure~\ref{fig:transits}). The amplitude of the noise is therefore in reasonable agreement with the predicted granulation signal, although the timescale is an order of magnitude longer than expected. Similar behavior was observed for NIRISS/SOSS observations of another M dwarf, LHS 1140, in \citet{Cadieux2024}: those authors measured a granulation timescale of 5 minutes and amplitude of 53~ppm. This amplitude is similar to the estimated amplitude of 32~ppm from the \citet{Gilliland2011} relations, but an order of magnitude longer than the predicted timescale of 40~s. However, this overestimation does not appear to be specific to M dwarfs: \citet{Coulombe2025} found the same effect in their analysis of NIRISS/SOSS observations of LTT~9779, a Sun-like star, measuring a granulation amplitude of 93 ppm and timescale of 21 minutes, in comparison to estimates of 70 ppm and 3 minutes. They argue that the incongruity is likely the result of the source of the correlated noise truly being a combination of granulation, mesogranulation, and supergranulation, which occur on different timescales.

\begin{figure}[t]
\centering
    \includegraphics[width=\columnwidth]{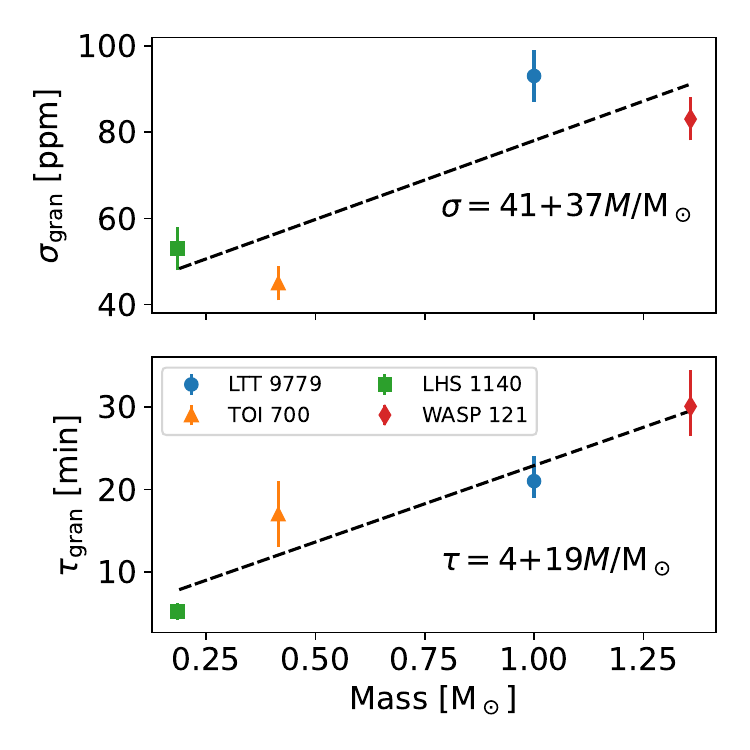}
    \caption{Amplitudes and timescales of the granulation-like signature observed in main-sequence stars using NIRISS/SOSS. The TOI~700 results are from this work, while LHS~1140 was studied in \citet{Cadieux2024}, LTT~9779 in \citet{Coulombe2025}, and WASP~121 in \citet{Splinter2025}. Although the sample size is small, the data suggest a trend of $\sigma_{\rm gran}$ and $\tau_{\rm gran}$ increasing with stellar mass. The equation of the best-fit line is provided on each plot.}
    \label{fig:granulation}
\end{figure}

To further investigate this issue, we can also consider the granulation signatures that have been observed in Kepler data. Equation 4 of \citet{Kallinger2014} reports an empirical relation linking $\tau_{\rm gran}$ to effective temperature and surface gravity based on a Kepler sample (which again does not contain any M dwarfs). Notably, their adopted solar value is larger than was used in the \citet{Gilliland2011} relation (375~s vs 220~s). The \citet{Kallinger2014} relation yields a predicted granulation timescale of roughly 7 minutes for all of LHS~1140, TOI~700, and LTT~9779, as they found that this quantity was primarily determined by surface gravity, with a very weak dependence on effective temperature. These values are in closer agreement with the timescales determined from the NIRISS/SOSS datasets, although the \citet{Kallinger2014} relation does not reproduce the diversity that appears to be present as a function of spectral type for these main-sequence dwarf stars; while the sample size is too small to draw robust conclusions, preliminary results suggest a linear trend (Figure~\ref{fig:granulation}). In addition to the three stars previously mentioned, Figure~\ref{fig:granulation} includes WASP 121, another stellar host with correlated noise in its NIRISS/SOSS white-light curve that \citet{Splinter2025} attribute to granulation. However, this star has an effective temperature of 6628$\pm$66~K \citep{Sing2024}, which overlaps within errors with recent estimates for the location of the Kraft Break \citep[6550$\pm$200~K; ][]{Beyer2024}, the temperature at which stars cease to have an outer convective envelope. It is thus unclear whether we expect granulation to occur in WASP 121.

Alternatively, instead of granulation, \citet{Holmberg2023} suggest that correlated noise in NIRISS/SOSS white-light curves may result from an imperfect 1/$f$ correction. To investigate this possibility, we rereduce our data using the three 1/$f$ correction methods in \texttt{exoTEDRF}: \texttt{scale-achromatic} (the default), \texttt{scale-achromatic-window}, and \texttt{solve}. While the \texttt{scale-achromatic-window} method (which only uses rows within 30 pixels of the trace to estimate the 1/$f$ level) is recommended to reduce residual red noise \citep{Radica2025}, we do not observe a difference in RMS error in our white-light curves based on our choice of 1/$f$ correction algorithm. We also explore the likelihood method for jump detection and ramp fitting from \citet{Brandt2024b, Brandt2024a}, which \citet{Carter2025} argue can reduce the noise in extracted NIRISS/SOSS white-light curves; however, we again do not find any improvement for our data set.  This insensitivity to reduction scheme supports the hypothesis that the correlated noise is astrophysical, although we cannot exclude the possibility that some of the excess noise stems from limitations in current best practices for NIRISS/SOSS data reduction. As illustrated by \citet{Kipping2026} for NIRSPEC, we still have long way to go to fully understand and correct all of the systematics in JWST data.

\subsubsection{Eccentricity}
\label{sec:ecc}
\citet{Gilbert2023} found that the eccentricities of all four planets were consistent with zero within the uncertainties of the TESS data. For this reason, we have assumed circular orbits in the fits we have previously discussed. However, given the enhanced photometric precision of our NIRISS/SOSS observations, we should reevaluate this assumption: our precision may be sufficient to resolve asymmetries in the light curve indicative of eccentricity \citep[e.g.,][]{Kipping2008} that TESS could not. We therefore relax our circular orbit assumption and repeat the previous fit, adopting the \citet{VanEylen2019} mixture-distribution eccentricity prior implemented in \texttt{exoplanet}. This prior describes the population of well-characterized small transiting planets from Kepler that are in multi-planet systems. Our resulting eccentricities for TOI 700 d and e are both consistent with zero, and we can place 95\% confidence upper limits of $e < 0.11$ for d and $e < 0.09$ for e when using the version of the fit that corrects the red noise with a GP. Without a GP, we achieve similar albeit modestly weaker results: $e < 0.15$ for d and $e < 0.12$ for e. Note that if the posterior was informed solely by the prior, we would find $e<0.25$; thus, the data support a low-eccentricity solution more strongly than the prior alone.  We therefore will continue to assume circular orbits for the joint fit in the following section.

\subsection{JWST/TESS/Spitzer joint fit}
\label{sec:joint}
We fit our JWST transits jointly with the 21 sectors of TESS observations analyzed in \citet{Gilbert2023}, as well as 16 additional sectors observed since the publication of that work (61--66, 68--69, 87--90, 93, 95--97). We also consider the two Spitzer \citep{Fazio2004} \hbox{transits of d from \citet{Rodriguez2020}.}

We again use \texttt{exoCTK} \citep{Bourque2021} to estimate quadratic limb-darkening parameters for TOI~700 in the TESS and Spitzer/IRAC2 bands, using normally distributed priors centered at these values and with $\sigma=0.1$. We use uninformative uniform priors for the radius ratio, transit duration, and impact parameters. We also include a normalization term and a white-noise jitter term for each time series, as well as a GP for each instrument that represents red noise (described in Section~\ref{sec:stellar}), all with broad, uninformative priors. For the times of transit and periods, we use normally distributed priors centered at the literature values and with $\sigma=0.01$, allowing these quantities to deviate if necessitated by the new data. For the stellar mass and radius, we again adopt priors of $M_*=0.417\pm0.021$M$_\odot$ and $R_*=0.419\pm0.021$R$_\odot$, as calculated in \citet{Gilbert2023} using the $K_s$-band relations from \citet{Mann2015}.

\begin{figure*}[h]
\centering
    \makebox[\textwidth][c]{%
    \includegraphics[width=1.07\textwidth]{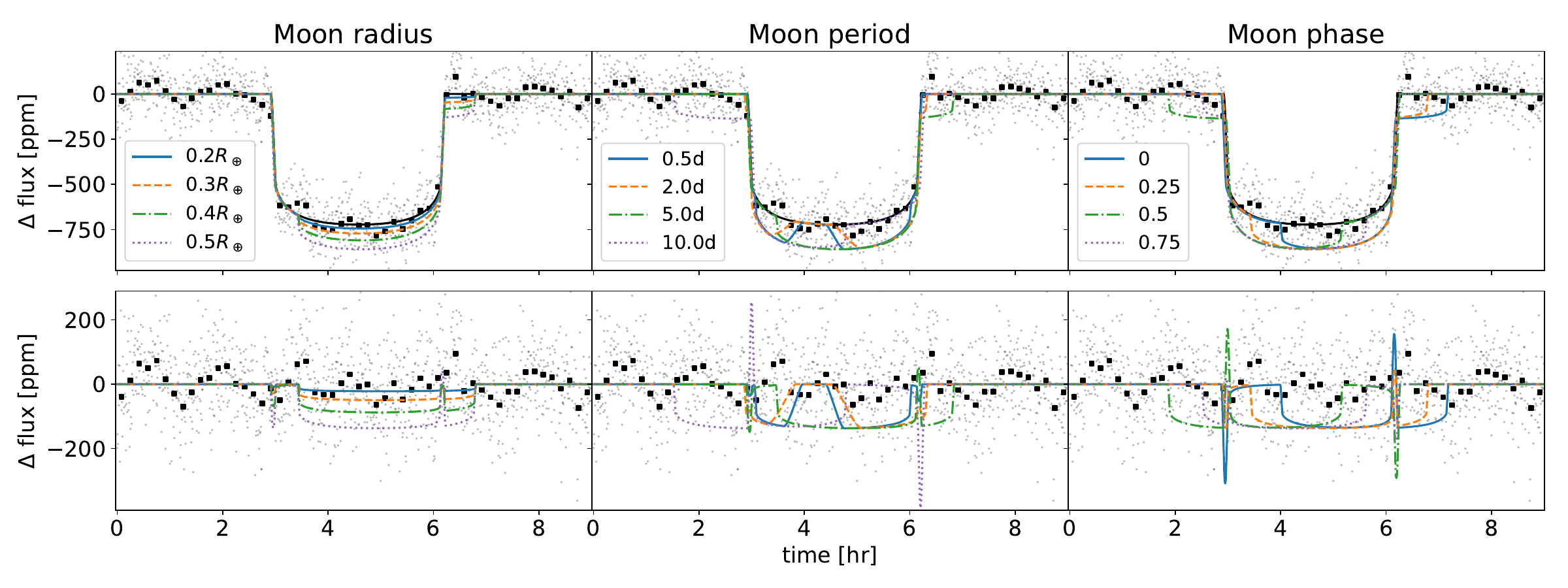}%
}
    \caption{We use the \texttt{pandora} package to illustrate the effects of varying moon parameters on the light curve. The black line in the upper panels shows a fiducial planet-only model that matches our observations of TOI~700~d, which we also plot to contextualize the amplitude of potential moon-induced changes. The colored lines illustrate a moon+planet model: from left to right, we vary the moon's radius, period, and phase. When not varying a given parameter, we adopt 0.5R$_\oplus$ in radius, 7~days in period, and 0.25 in phase. The lower panels show the residuals after subtracting the fiducial model. The presence of a moon can influence the transit depth and time of transit (effects that we cannot easily distinguish from the planet-only model in a single transit), as well as the transit shape (effects that are more readily distinguishable in a single epoch of observation).}
    \label{fig:sim}
    \vspace{1cm}
    \includegraphics[width=1.\textwidth]{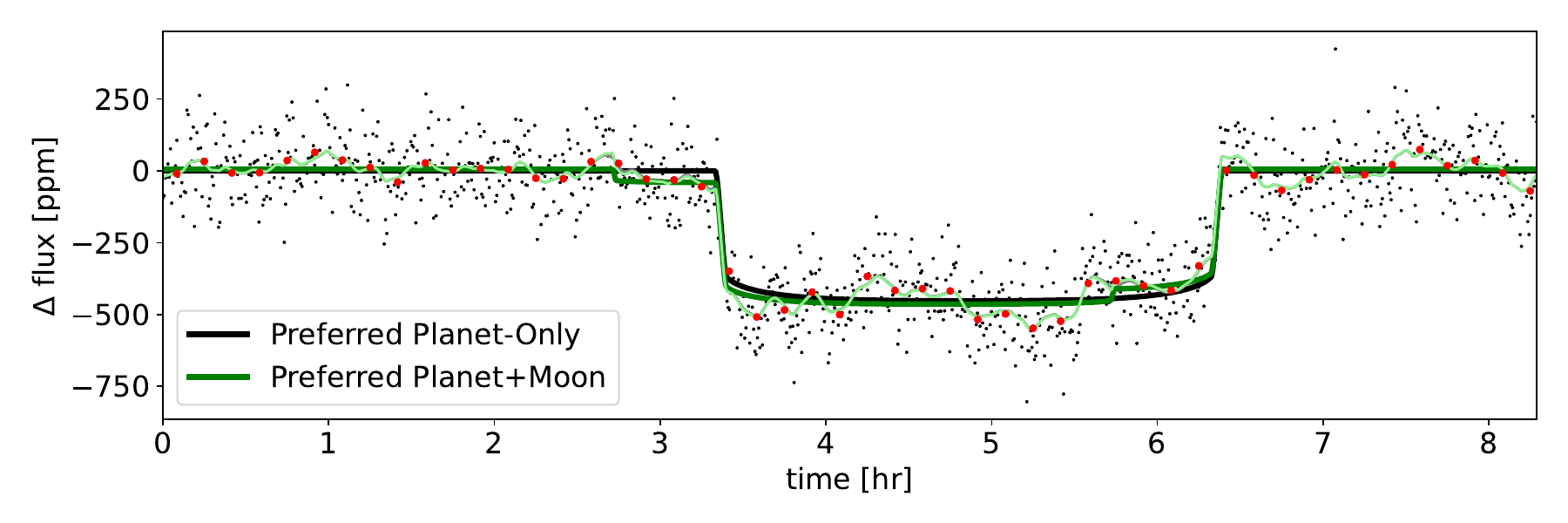}
    \caption{For TOI~700~e, the best-fitting planet-only model is shown in black and the planet+moon model is shown in green. The modelled moon has a period of 5.8 days and a radius of 0.31R$_\oplus$. The models including the GPs are shown in lighter shades of gray/green; the GP models are nearly identical between the two solutions. The planet+moon model is marginally preferred, with a Bayes difference of $\Delta\log(Z)=0.14$. The planet is slightly smaller in the planet+moon case, thereby producing the same depth of transit during the times when both planet and moon are crossing in front of the star.}
    \label{fig:toi700-ganymede}
\end{figure*}

To determine best-fitting parameters and uncertainties, we sample the posterior of our model using a Markov-Chain Monte Carlo. Starting from the maximum \textit{a posteriori} solution, we use the modified \texttt{PyMC3} \citep{Salvatier2016} sampler implemented in \texttt{exoplanet} to sample four chains each with a 2000-draw burn-in and 3000 draws. We use an initial acceptance fraction of 0.5, a target acceptance fraction of 0.95, and 100 regularization steps. We find that the sampler properly converges, as evidenced by Gelman–Rubin statistics \citep{Gelman1992} near 1 for all parameters. Our results are given in Table~\ref{tab:results}; compared to the previous best-fit solution from \citet{Gilbert2023}, we are able to improve the precision in the radii of d and e by factors of 2--3, and the precision in the periods by factors of 5--20.

\subsection{Search for moons}
\label{sec:search}
\subsubsection{\texttt{pandora} model}
To search our datasets for moons, we use the open-source exomoon transit code \texttt{pandora} \citep{Hippke2022}. This code performs analytical photodynamical modelling of both planet and moon, allowing it to incorporate effects such as planet--moon eclipses and barycentric motion of the planet+moon system.

Unlike in our previous analysis, our goal here is not to accurately determine the uncertainties in each planetary parameter and the degeneracies between parameters; rather, our focus is on the range of plausible transit shapes allowed for a planet-only model, and how these transit shapes compare to the plausible range of planet+moon models. To this end, we consider eight parameters for our planet-only model: the stellar density, the planet-star radius ratio, the impact parameter, the time of transit, the two parameters of a quadratic limb darkening law, a normalization term for the light curve, and a white-noise jitter term. Because we are only considering a single transit, we fix the period to the value obtained in the previous joint fit for the purposes of this analysis. We also fix the value of the stellar radius. We note that it would be inadvisable to fix the transit duration and depth to the values found in the joint fit: these parameters are mostly constrained by the JWST data and hence would be biased if a moon exists in the system. We therefore adopt uninformative uniform priors for all other parameters except the stellar density and limb darkening. For density, we adopt a normally distributed prior with $\rho_*=8.5\pm0.6$~gcm$^{-3}$, based on the results of our previous joint fit (this value is informed by the ensemble of all transits of b, c, d, and e, and hence is robust even if a moon is present in one of the JWST transits). For limb darkening, we use the same \texttt{exoCTK}-determined prior from the previous sections. We also include the same \texttt{celerite2} GP as before, with the amplitude and length scale hyperparameters fixed to their best-fit values from the JWST-only analysis.

Our planet+moon model contains three additional parameters: the moon's radius ratio, period, and phase. In Figure~\ref{fig:sim}, we illustrate the impact of changing these parameters on the resultant light curve. For the purposes of this analysis, we assume the moon and planet are coplanar; a preference towards such alignments is expected given tidal considerations. We otherwise adopt uninformative uniform priors for our moon parameters. In the case of period, we use a range of 0.1--10 days. Assuming a terrestrial mass--radius relationship, the Roche limit is roughly 0.26 days, and 0.4895 times the Hill radius is 5.5 days for e and 7.4 days for d; our period prior is thus conservative relative to the parameter space in which we expect moons to exist.

\begin{figure*}[t]
\centering
    \includegraphics[width=0.49\textwidth]{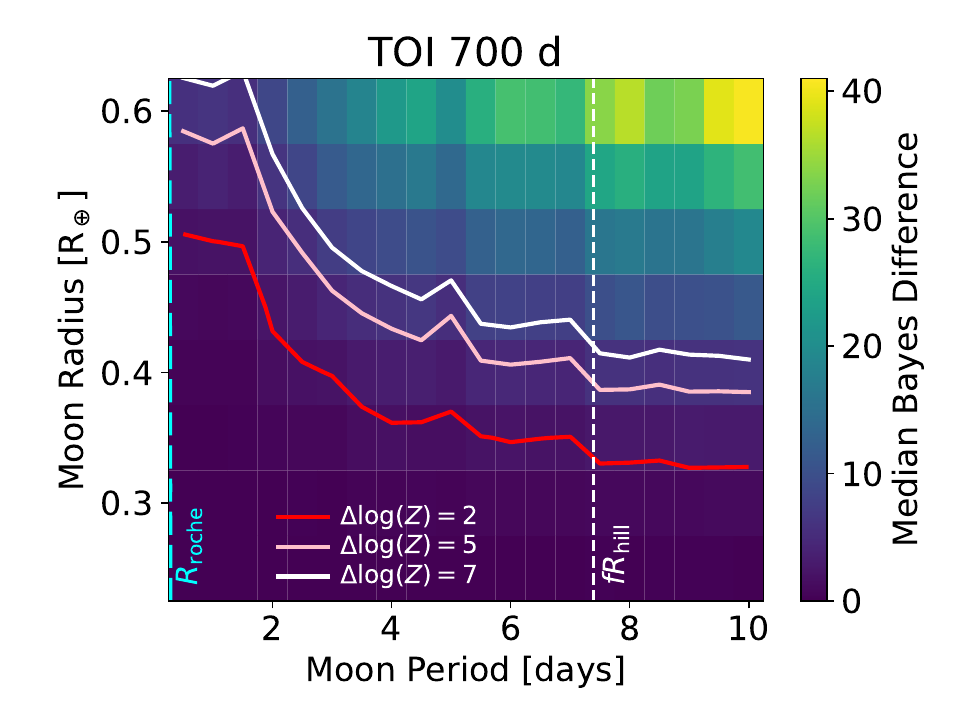}
    \includegraphics[width=0.49\textwidth]{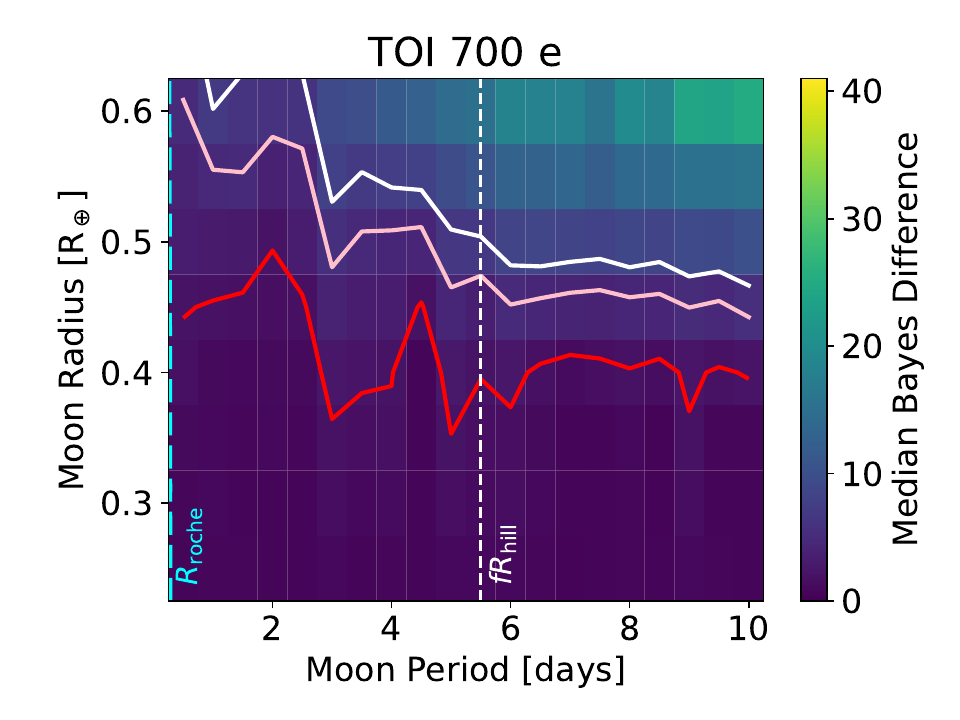}
        \includegraphics[width=0.49\textwidth]{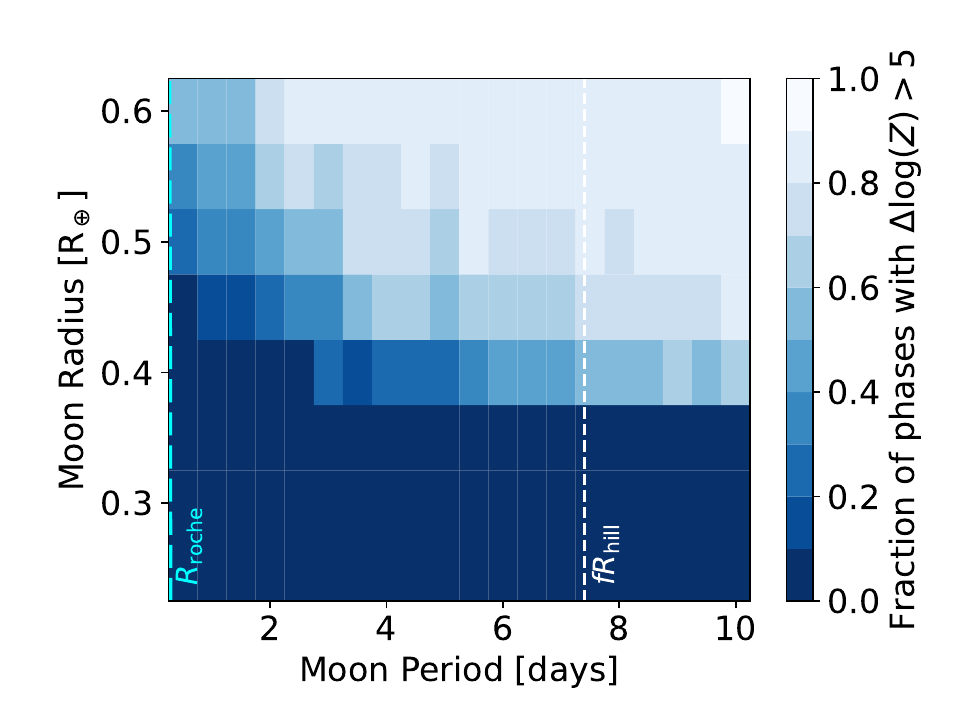}
    \includegraphics[width=0.49\textwidth]{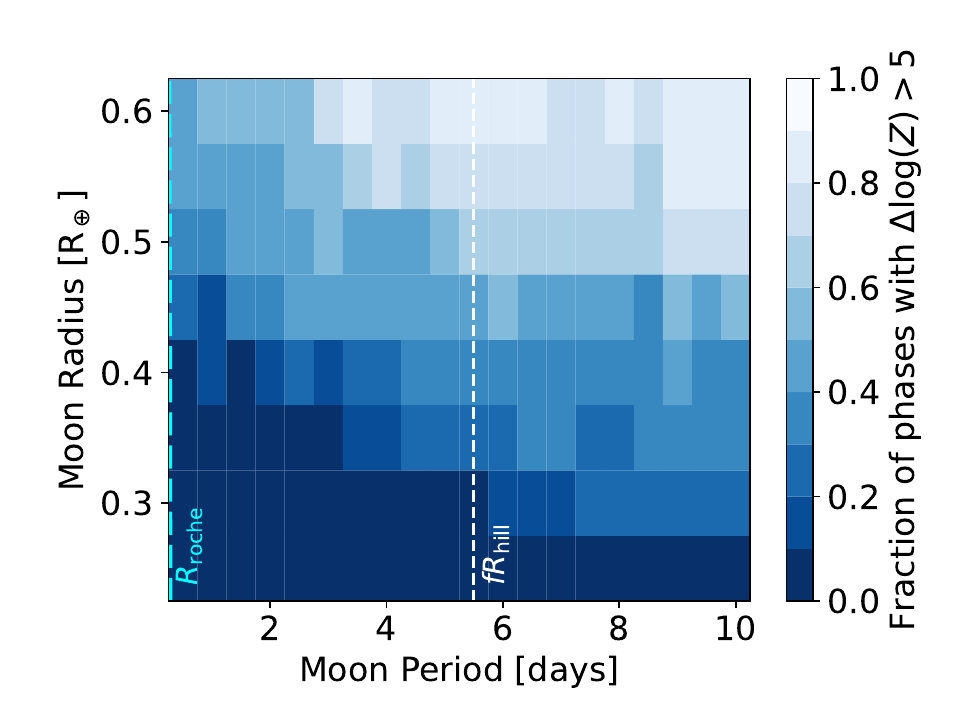}
    \caption{The left plots show the results of our injection and recovery test into our JWST observation of TOI 700 d, and the right for e. The Roche limit and $f$ times the Hill radius are noted, where $f=0.4895$ \citep{Domingos2006}; these lines bound the region in which a prograde satellite is expected to be dynamically stable. In the upper panels, the heatmap provides the Bayes difference between the planet-only and planet+moon models, taking the median of the results across the 100 injections. The red solid line indicates the $\Delta\log(Z)=2$ contour, the pink $\Delta\log(Z)=5$, and the white $\Delta\log(Z)=7$ (corresponding to 7.4:1, 148:1, and 1097:1 odds in favor of the planet+moon model, respectively). In the lower panels, we plot the fraction of the 100 phases that yield $\Delta\log(Z)>5$. These plots illustrate that there are certain phases at which it is particularly challenging to differentiate a moon from a larger planet-only solution given the noise present in our data. It is also challenging to detect a moon whose period is less than 2 days, as \hbox{the shape changes at ingress and egress become more subtle than the noise in our light curves.}}
    \label{fig:injection}
\end{figure*}

We derive posterior probability distributions and the Bayesian evidence with the nested sampling Monte Carlo algorithm \texttt{MLFriends} \citep{Buchner2016, Buchner2019} using the \texttt{UltraNest} package \citep{Buchner2021}. We base our implementation on the recommended workflow provided in the \texttt{pandora} documentation, as the developers investigated the various \texttt{UltraNest} sampler options and found that the \texttt{RegionSliceSampler} yielded optimal performance. We set a target evidence uncertainty of $\Delta\log(Z)=0.2$ and a minimum of 600 live points.

We consider the TOI~700~d and e light curves separately, fitting each with the planet-only and planet+moon models and calculating the Bayes difference. For TOI~700~d, the best-fitting planet+moon model is not preferred over the the best-fitting planet-only model. For TOI~700~e, the best-fitting planet+moon model is slightly preferred with a Bayes difference of $\Delta\log(Z)=0.14$; i.e., a Bayes factor of e$^{0.14}=1.15$. This best-fitting moon has a period of 5.8 days and a radius of 0.31R$_\oplus$ (slightly larger than Luna). As outlined in \citet{Thorngren2026}, there have historically been misconceptions in the exoplanet field regarding the appropriate interpretation of the statistical significance of a given Bayes factor. Following Equation 4 of \citet{Thorngren2026}, this Bayes difference yields a null probability of 1/(1+e$^{0.14})=$0.46. Or put otherwise, it represents 1.15:1 odds in favor of the planet+moon model. This is not statistically significant.

\begin{figure*}[t]
\centering
    \includegraphics[width=0.49\textwidth]{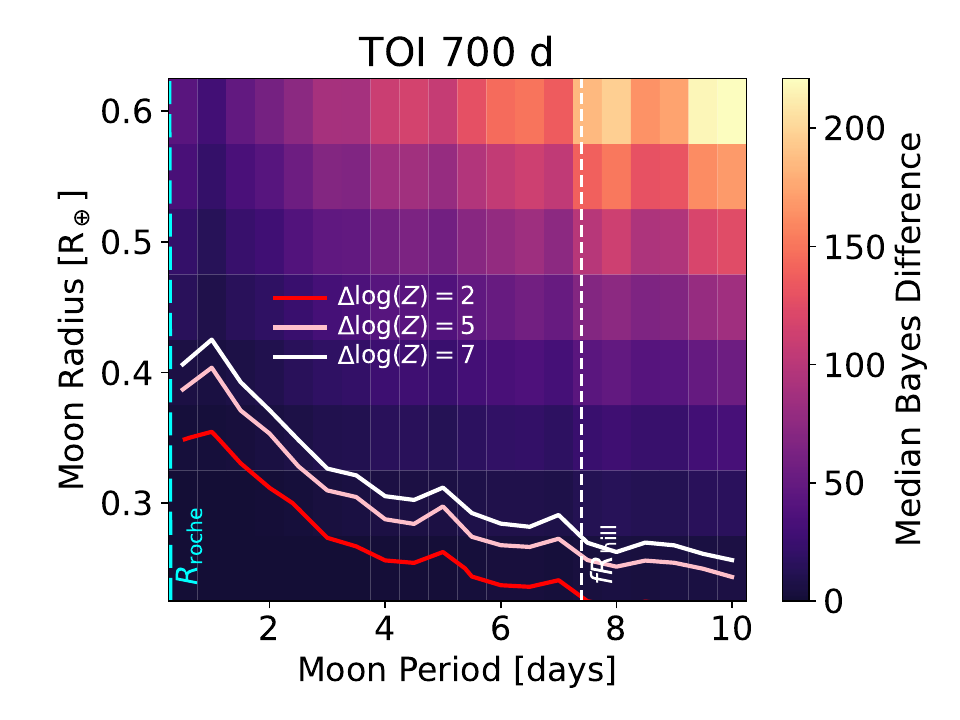}
    \includegraphics[width=0.49\textwidth]{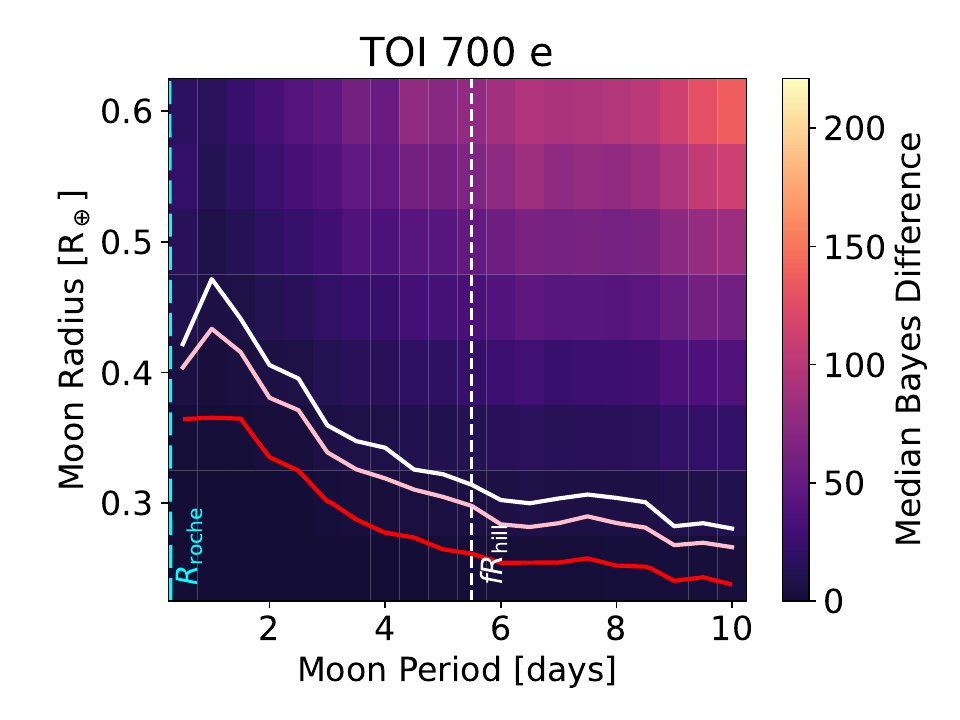}
    \includegraphics[width=0.49\textwidth]{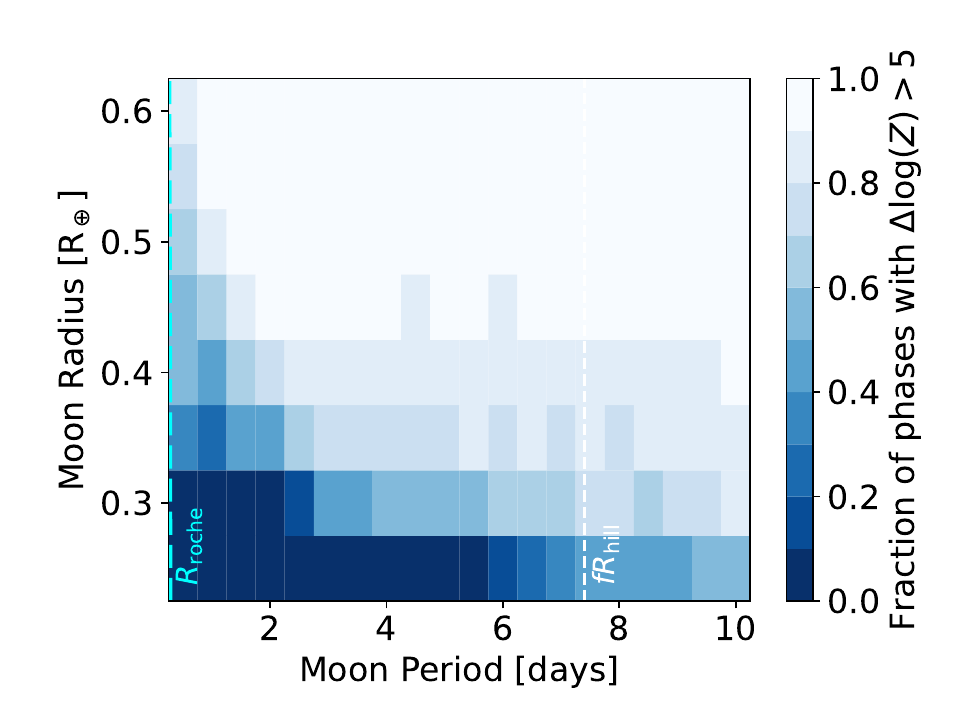}
    \includegraphics[width=0.49\textwidth]{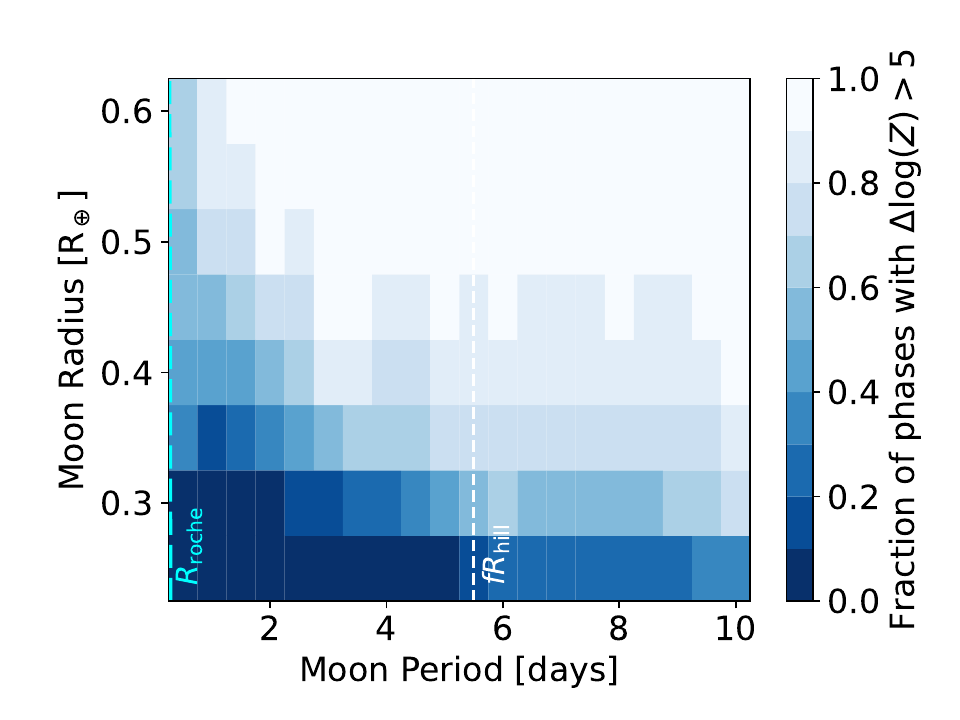}
    \caption{The same as Figure~\ref{fig:injection}, but for the idealized case where we have removed the correlated noise using the GP fit from Figure~\ref{fig:transits} before conducting our injection and recovery test. These figures thus represent the sensitivity to moons that is possible with the current data if future analyses are able to mitigate this noise source and recover the photon-noise limit.}
    \label{fig:injection-ideal}
\end{figure*}

From inspection of these models in Figure~\ref{fig:toi700-ganymede}, it is clear that the presence of correlated noise in our JWST time series is a severe impediment to our ability to detect moons. The differences between the planet-only and planet+moon light curves occur on a timescale comparable to the GP length scale, and one is therefore unlikely to be convinced that the tentative signal is truly caused by a moon; the GP has sufficient flexibility to subsume the differences between the two models, such that the two model+GP fits (plotted in gray and light green) are virtually identical. The correlated noise thus sets our noise floor: while it is nominally `corrected' by the GP, in practice, its presence means that our noise floor is a factor of four larger than it would be in the photon-noise limit, and as a result, moons that are even larger than Luna can be hiding in our data without producing a statistically robust signal.

\subsubsection{Sensitivity analysis}
Of course, we nonetheless want to quantify what moons can be ruled out by our observations. We therefore conduct an injection and recovery test. We consider twenty moon periods spaced in 0.5-day increments from 0.5 to 10 days and eight moon radii spaced in 0.05-R$_\oplus$ increments from 0.25 to 0.60R$_\oplus$. We sample 100 phases drawn from a random uniform distribution between 0 and 1, injecting these 16000 simulated moons into our real JWST light curves of each planet. We analyze the modified light curves in the same manner described above and calculate the Bayes difference between the best-fitting planet-only model and planet+moon model. Note that the best-fitting planet-only model is different in each injection, as the presence of the simulated moon can affect the planet-only fit (for example, by biasing towards larger planet radii).

In the upper panels of Figure~\ref{fig:injection}, we plot the Bayes difference as a function of the moon's period and radius, taking a median across the 100 injections. In red, we indicate the contour where the median Bayes difference is 2; i.e., the planet+moon model is favored with 7.4:1 odds and the null probability is 0.12. In pink, we show a median Bayes difference of 5 (148:1 odds and a null probability of 0.0067), and 7 in white (1097:1 odds; null probability of 0.00091). Our sensitivity generally improves for larger moons on longer orbits, as such moons create deeper transits and produce shape effects that are more readily distinguishable from a larger planet-only solution. There are some phases at which it is more difficult to distinguish the moon given our noise characteristics, as illustrated in the lower panels that show the fraction of phases with $\Delta\log(Z)>5$. These results highlight the usefulness of obtaining multiple transits per planet, as it would be increasingly unlikely to obtain repeat \hbox{observations during uncommon, unfavorable phases.}

From this analysis, we conclude that our observations are generally only sensitive to moons larger than Ganymede (the largest moon in our solar system; radius 0.41R$_\oplus$) on orbits longer than 2 days. The dominant factor limiting our sensitivity is the presence of correlated noise on a roughly 16 minute timescale that inflates our errors to four times the photon-noise limit.

We can also quantify the information content that exists in these JWST data if future analyses are able to reduce the impact of correlated noise and recover the photon-noise limit (through either an improved understanding of the relevant stellar physics or JWST systematics, depending on the noise's true origin). To this end, we repeat our injection and recovery analysis, but replace the observed JWST white-light curve with a version that has had the GP fit removed. We then do not include a GP in our \texttt{pandora} fits. These results are shown in Figure~\ref{fig:injection-ideal}. We find that we are indeed sensitive to Luna-analog moons around TOI~700~d in this idealized scenario, and slightly larger moons for e; future work to mitigate the impact of correlated noise is thus essential to unlock this exciting science case.

\section{Summary and Conclusions}
\label{sec:conclusion}
Using the NIRISS/SOSS mode of JWST, we collect a transit of each of the M-dwarf habitable-zone rocky planets TOI~700~d and e. We analyze these data jointly with previous observations from TESS and Spitzer, significantly improving the orbital ephemerides and the inferred planetary radii as a result of JWST's exquisite photometric precision. Our best-fit parameters and uncertainties are given in Table~\ref{tab:results}. To highlight some key gains, we refine the radius error of d from 5.3\% to 2.5\% and of e from 8.6\% to 2.8\%, and we improve the period uncertainties by an order of magnitude.

The limiting noise source in our observations is a time-correlated signal with a characteristic timescale of 16$\pm$4~minutes and amplitude of 46$\pm$4~ppm. This signal is apparent in both the transit of d and e (Figure~\ref{fig:transits}) and is not strongly chromatic (Figure~\ref{fig:chromatic}). Similar signals have been previously identified in JWST observations of other stars and attributed to stellar granulation \citep{Cadieux2024, Coulombe2025}. We show that the timescales and amplitudes of these signals vary with host star mass (Figure~\ref{fig:granulation}), supporting the granulation hypothesis. However, granulation was not predicted to be a dominant effect for such red stars and at such red wavelengths \citep[e.g.,][]{Sarkar2018}, and we do not observe the expected chromaticity despite broad wavelength coverage (0.60--2.83\textmu m). We also investigate the potential impacts of known instrumental systematics and are unable to identify an instrumental explanation for the correlated noise, although we cannot rule out a yet-unknown instrumental effect. Practically speaking, this noise source limits our precision to a factor of 4 above the photon-noise limit in 10-minute bins.

The goal of our investigation was to search for exomoons around TOI~700~d and e, with the expectation that our observations would provide sensitivity to a Luna-analog moon (0.27R$_\oplus$). The presence of correlated noise prevents us from achieving that objective, and we find that we are instead sensitive mainly to moons larger than Ganymede ($>$0.4R$_\oplus$) on periods longer than 2 days (Figure~\ref{fig:injection}). We illustrate that if the correlated noise could be corrected or mitigated, our observations would be sufficient to place informative constraints on the occurrence of Luna-sized moons (Figure~\ref{fig:injection-ideal}). These results thus highlight a pressing need to better understand this correlated noise source, be it stellar or instrumental---not just to permit detection of exomoons that are analogous to the natural satellites of our solar system, but to enable all science cases that require 10~ppm precision in JWST white-light curves.

\section*{Acknowledgments}
%TC:ignore
E.P.\ is supported by a Juan Carlos Torres Postdoctoral Fellowship at the Massachusetts Institute of Technology and A.V.\ is supported in part by a Sloan Research Fellowship. We thank Benjamin Rackham for a helpful discussion about stellar granulation and the anonymous referee for their valuable comments.

This work is based on observations made with the NASA/ESA/CSA James Webb Space Telescope. The data were obtained from the Mikulski Archive for Space Telescopes (MAST) at the Space Telescope Science Institute, which is operated by the Association of Universities for Research in Astronomy, Inc., under NASA contract NAS 5-03127 for JWST. These observations are associated with program \#6193. Support for program \#6193 was provided by NASA through a grant from the Space Telescope Science Institute, which is operated by the Association of Universities for Research in Astronomy, Inc., under NASA contract NAS 5-03127. The specific observations analyzed can be accessed via\dataset[doi: 10.17909/gq8f-6795]{https://doi.org/10.17909/gq8f-6795}.

This paper includes data collected by the TESS mission, which are publicly available from the MAST. Funding for the TESS mission is provided by the NASA's Science Mission Directorate. It also includes observations made with the Spitzer Space Telescope, which was operated by the Jet Propulsion Laboratory, California Institute of Technology under a contract with NASA.

The authors acknowledge the MIT Office of Research Computing and Data for providing high performance computing resources that have contributed to the research results reported within this paper.


\facilities{JWST, TESS, Spitzer}
\software{\texttt{exoTEDRF} \citep{Radica2024}, \texttt{lightkurve} \citep{lightkurve2018}, \texttt{Matplotlib} \citep{Hunter2007}, \texttt{NumPy} \citep{Harris2020}, \texttt{pandas} \citep{Reback2021}, \texttt{pandora} \citep{Hippke2022}, \texttt{scikit-learn} \citep{Pedregosa2011}, \texttt{SciPy} \citep{Scipy2020}, \texttt{UltraNest} \citep{Buchner2021}, as well as \texttt{exoplanet} \citep{ForemanMackey2021,
ForemanMackey2021a} and its dependencies \citep{ForemanMackey2017,
ForemanMackey2018, Agol2020, Kumar2019,
Astropy2013, Astropy2018, Astropy2022, Kipping2013,
Luger2019, Salvatier2016, TDT2016}.}
%TC:endignore

%% For this sample we use BibTeX plus aasjournalv7.bst to generate the

\bibliography{moon}{}
\bibliographystyle{aa_url}



\begin{deluxetable}{l l c c c}[t]
\tablecaption{Results of JWST/TESS/Spitzer joint fit for TOI~700 system parameters\label{tab:results}}
\tablewidth{0pt}
\tablehead{
\colhead{Parameter} & \colhead{Description (Units)}  & \colhead{Median} & \colhead{+1$\sigma$}  & \colhead{-1$\sigma$}
}
\startdata
\textbf{TOI 700}\\
$R_*$\dotfill &Stellar radius (R$_\odot$)\dotfill&0.4104&0.0093&0.0069\\
$M_*$\dotfill &Stellar mass (M$_\odot$)\dotfill&0.417&0.020&0.020\\
$\rho_*$\dotfill &Stellar density (gcm$^{-3}$)\dotfill&8.50&0.52&0.64\\
\textbf{Limb darkening}\\
$u_{1,\rm{j}}$\dotfill &Linear, NIRISS/SOSS \dotfill&0.162&0.066&0.067\\
$u_{2,\rm{j}}$\dotfill &Quadratic, NIRISS/SOSS \dotfill&0.269&0.081&0.082\\
$u_{1,\rm{t}}$\dotfill &Linear, TESS \dotfill&0.252&0.080&0.081\\
$u_{2,\rm{t}}$\dotfill &Quadratic, TESS \dotfill&0.305&0.090&0.090\\
$u_{1,\rm{s}}$\dotfill &Linear, IRAC2 \dotfill&0.055&0.062&0.039\\
$u_{2,\rm{s}}$\dotfill &Quadratic, IRAC2 \dotfill&0.092&0.089&0.075\\
\textbf{TOI 700 b}\\
$P$\dotfill &Period (days)\dotfill&9.977220&0.000012&0.000012\\
$a$\dotfill &Semi-major axis (AU)\dotfill&0.0678&0.0011&0.0011\\
$T_0$\dotfill & Time of conjunction (BJD)\dotfill&2458880.0994&0.0012&0.0013\\
$T_{14}$\dotfill &Transit duration (days)\dotfill&0.0887&0.0019&0.0021\\
$R_{P}$\dotfill &Radius (R$_\oplus$)\dotfill&0.963&0.037&0.034\\
$R_{P}$/$R_{*}$\dotfill & Radius ratio \dotfill&0.02150&0.00062&0.00065\\
$i$\dotfill &Inclination (degrees)\dotfill&89.63&0.23&0.21\\
$b$\dotfill &Impact parameter \dotfill&0.23&0.12&0.14\\
\textbf{TOI 700 c}\\
$P$\dotfill &Period (days)\dotfill&16.0511039&0.0000056&0.0000055\\
$a$\dotfill &Semi-major axis (AU)\dotfill&0.0931&0.0015&0.0015\\
$T_0$\dotfill & Time of conjunction (BJD)\dotfill&2458821.62190&0.00041&0.00043\\
$T_{14}$\dotfill &Transit duration (days)\dotfill&0.05828&0.00106&0.00096\\
$R_{P}$\dotfill &Radius (R$_\oplus$)\dotfill&2.535&0.084&0.073\\
$R_{P}$/$R_{*}$\dotfill & Radius ratio \dotfill&0.0566&0.0012&0.0011\\
$i$\dotfill &Inclination (degrees)\dotfill&88.942&0.032&0.036\\
$b$\dotfill &Impact parameter \dotfill&0.8992&0.0073&0.0074\\
\textbf{TOI 700 e}\\
$P$\dotfill &Period (days)\dotfill&27.810124&0.000071&0.000087\\
$a$\dotfill &Semi-major axis (AU)\dotfill&0.1336&0.0020&0.0020\\
$T_0$\dotfill & Time of conjunction (BJD)\dotfill&2460772.46343&0.00024&0.00024\\
$T_{14}$\dotfill &Transit duration (days)\dotfill&0.12747&0.00050&0.00048\\
$R_{P}$\dotfill &Radius (R$_\oplus$)\dotfill&0.919&0.028&0.024\\
$R_{P}$/$R_{*}$\dotfill & Radius ratio \dotfill&0.02053&0.00036&0.00039\\
$i$\dotfill &Inclination (degrees)\dotfill&89.890&0.076&0.099\\
$b$\dotfill &Impact parameter \dotfill&0.134&0.116&0.093\\
\textbf{TOI 700 d}\\
$P$\dotfill &Period (days)\dotfill&37.423457&0.000018&0.000017\\
$a$\dotfill &Semi-major axis (AU)\dotfill&0.1642&0.0025&0.0027\\
$T_0$\dotfill & Time of conjunction (BJD)\dotfill&2460763.01612&0.00021&0.00023\\
$T_{14}$\dotfill &Transit duration (days)\dotfill&0.13762&0.00046&0.00046\\
$R_{P}$\dotfill &Radius (R$_\oplus$)\dotfill&1.145&0.032&0.026\\
$R_{P}$/$R_{*}$\dotfill & Radius ratio \dotfill&0.02559&0.00031&0.00032\\
$i$\dotfill &Inclination (degrees)\dotfill&89.830&0.086&0.068\\
$b$\dotfill &Impact parameter \dotfill&0.255&0.093&0.126
\enddata
\end{deluxetable}

\end{document}